\documentclass[conference]{IEEEtran}
\IEEEoverridecommandlockouts
\usepackage{cite}
\usepackage{amsmath,amssymb,amsfonts}
\usepackage{multirow}

\usepackage{graphicx}
\usepackage{textcomp}
\usepackage{xcolor}
\usepackage{comment}
\usepackage{booktabs}
\usepackage{graphicx}
\usepackage{tabularx}
\usepackage{tablefootnote}
\usepackage{array}
\newcommand{\spheading}[2][8em]{
  \rotatebox{90}{\parbox{#1}{\raggedright #2}}}
\usepackage{wasysym}
\usepackage{tablefootnote}
\usepackage{soul}

\usepackage{hyperref}
\usepackage{pifont}
\usepackage{tikz}
\usetikzlibrary{shapes.geometric, arrows}
\newcommand{\xmark}{\ding{55}}%
\newcommand{\cmark}{\ding{51}}%
\usepackage[caption=false]{subfig}
\usepackage{threeparttable}
\usepackage{lineno} 
\usepackage{multicol}
\usepackage{blindtext}
\usepackage{lscape}
\usepackage{tikz}
\usepackage{tikz-qtree}
\usetikzlibrary{patterns,arrows,positioning,calc,intersections,trees, chains, quotes, shapes.misc,
decorations.pathmorphing,positioning,decorations.pathreplacing,patterns,shapes.geometric, shapes.multipart,arrows.meta}
\def\BibTeX{{\rm B\kern-.05em{\sc i\kern-.025em b}\kern-.08em
    T\kern-.1667em\lower.7ex\hbox{E}\kern-.125emX}}
    
\begin{document}

\title{Blockchain Simulators: A Systematic Mapping Study}


\author{\IEEEauthorblockN{Adel Albshri\IEEEauthorrefmark{1}\IEEEauthorrefmark{2},Ali Alzubaidi\IEEEauthorrefmark{3}\IEEEauthorrefmark{4}, Bakri Awaji,\IEEEauthorrefmark{5}\IEEEauthorrefmark{6}  and Ellis Solaiman\IEEEauthorrefmark{7}}

\IEEEauthorblockA{\IEEEauthorrefmark{1}Newcastle University, School of Computing, UK, Email: a.albshri2@ncl.ac.uk}

\IEEEauthorblockA{\IEEEauthorrefmark{2}University of Jeddah, Saudi Arabia, 
Email: amalbeshri@uj.edu.sa}

\IEEEauthorblockA{\IEEEauthorrefmark{3}Umm Al-Qura University, Saudi Arabia, 
Email: aakzubaidi@uqu.edu.sa}

\IEEEauthorblockA{\IEEEauthorrefmark{4}Newcastle University, School of Computing, UK, Email: aakzubaidi@IEEE.org}

\IEEEauthorblockA{\IEEEauthorrefmark{5}Newcastle University, School of Computing, UK, Email: b.h.m.awaji2@ncl.ac.uk}

\IEEEauthorblockA{\IEEEauthorrefmark{6}Najran University, Saudi Arabia, 
Email: balawaji@nu.edu.sa}

\IEEEauthorblockA{\IEEEauthorrefmark{7}Newcastle University, School of Computing, UK, Email: ellis.solaiman@newcastle.ac.uk}}

\maketitle
\thispagestyle{empty}
\pagestyle{empty}
\begin{abstract}
Recently, distributed ledger technologies like blockchain have been proliferating and have attracted interest from the academic community, government, and industry. A wide range of blockchain solutions has been introduced, such as Bitcoin, Ethereum, and Hyperledger technologies in the literature. However, tools for evaluating these solutions and their applications are still lacking, limiting the exploration of their potentiality and associated challenges/limitations. That is, experimenting with real blockchain networks usually requires a solid budget; and thus, sophisticated blockchain simulators can facilitate designing and evaluating solutions before the actual implementation stage. The quality of such simulators depends on several factors such as usability, reliability, provided capabilities, and supported features. This paper aims to provide a systemic mapping review of blockchain simulators focusing on these quality factors. This paper also sheds light on the configuration parameters (inputs) and produced metrics (outputs) supported by each simulator. Furthermore, it investigates which metrics supported by each simulator are scientifically validated/evaluated. Moreover, code quality comparison is carried out to assess the source code of the covered simulators. The results reveal that no simulator fully covers the wide operational range of features and capabilities of existing blockchain technologies. However, several promising efforts exist in the domain of blockchain simulation with interesting and useful features. Finally, we discuss the subject of blockchain simulation and provide our insight into the matter.
\end{abstract}


\section{Introduction}
Traditionally, transactions and exchanges between parties have typically been carried out within a centralised structure, which requires the contribution of a third party (e.g. a bank). The hurdle is that this manner of transaction relies mainly on the third party in that if the party encounters a failure, the system completely stops. This problem is commonly known as a single point of failure (SPOF) \cite{Kumari2020}. What is more, high fees are often associated with third parties. Blockchain has arisen to handle these issues (both SPOF and high fees) by permitting nodes (parties) to associate with one another in a decentralized (aka distributed) way without the contribution of a third party. This is why this technology has gained researchers' attention and enthusiasm in recent years. Formally, blockchain uses a distributed/shared database that logs all the executed transactions within a network \cite{Abu-Elezz2020}. In other words, through the use of a distributed ledger, blockchain makes all transactions available and verifiable by all nodes. Interestingly, the involved nodes are transparent to the chain state update. In particular, thanks to the decentralized nature, nodes can transparently view all transactions occurring at a given time. Each node has its own copy of the chain that is updated following every newly confirmed block.  

In the beginning, blockchain was designed for handling the exchange of a digital currency – referred to as Bitcoin \cite{Nakamoto2008} – in a peer-to-peer network. Following the success of Bitcoin technology, a number of other blockchain solutions emerged, such as Ethereum in July 2015, and Hyperledger in December 2015. Since then, these solutions 
have been applied to various application domains, such as the Internet of Things (IoT) \cite{Umar2021}. Specifically, Bitcoin networks offered money-related exchanges through the utilisation of the eponymous tokens: bitcoins. The tokens subsequently developed an immense financial worth \cite{Chen2021}. Other blockchains like Ethereum \cite{wood2014ethereum} permitted code to be executed within blockchain system, granting flexibility to the transactions and exchanges, commonly known as smart contracts \cite{Omar2021}. By and large, the reason behind the success of blockchain technology is its wide range of merits. Firstly, blockchain creates immutable ledgers, which by nature are unable to be changed or altered. Once a transaction is created and registered, it cannot be altered \cite{Yiu2021}. Secondly, a vital characteristic of blockchain is its reliance on decentralized control, in which the resources of all nodes involved are used. This overcomes the issue of SPOF \cite{Omar2021}. Thirdly, blockchain protects the identity of users in an efficient manner. Fourthly, there is stronger security in blockchain technology due to the mitigation of the SPOF issue\cite{Andola2021}. Finally, blockchain enables participating nodes to collaboratively process transactions in a timely manner \cite{Xu2021}.

From a practical perspective, when it comes to applying a system – and blockchain is no exception – faults in real world application may result in major losses; for example, in cost, safety, resources, and environmental issues \cite{Hunhevicz2020}. In order to limit possible faults and unexpected failures, and in order to identify bottlenecks, simulations are commonly employed, using various design setups prior to the implementation of the actual design, or when making amendments to existing systems. Similar to emulation, simulation is of high value when tackling complicated tasks in a complicated environment \cite{Smetanin2020}.

In light of the need for a comprehensive analysis of current blockchain simulators, this paper intends to examine existing blockchain simulators, highlight their features, and identify possible challenges. To accomplish this task, a systematic mapping study was chosen as the methodology, in line with the systematic mapping process shown in \cite{Petersen2008}. Following this, related papers are scanned from scientific databases, and a map of existing blockchain-related systems has been subsequently developed. This map provides structured data and an in-depth insight in the domain of blockchain simulators.

The remainder of this paper is organised as follows: Section \ref{background} gives a background on blockchain technology, and highlights the need for blockchain simulation. Section \ref{related} reviews the literature with regard to similar systematic studies. The research methodology is given in Section \ref{methodology}. Section \ref{result} outlines the results of the systemic mapping. Section \ref{discussion} provides a detailed discussion about the current simulators and their limits. Finally, Section \ref{conclusion} concludes the paper.

\section{Preliminaries}
\label{background}
\subsection{Overview of Blockchain Technology}
Blockchain is a peer to peer (P2P) distributed network, which securely and immutably records all  transactions processed by participating nodes in the network \cite{Bhushan2021}. Blockchain diminishes the need for a trusted third party, as nodes share data directly (usually in the form of transactions) \cite{Miyachi2021}. Transactions are organised and ordered into blocks that are identified using cryptographic hashes where, each block points to the previous block; hence the name "blockchain".

Once a block is considered valid across blockchain network, it can never be altered or modified (immutable), and thus the contained transactions cannot be reverted. This acts as a firewall against the double-spending problem through achieving transaction integrity \cite{Efanov2018}. 

Generally, blockchain networks today are classified into three categories: public, private, and hybrid \cite{W2018}.
\begin{itemize}
    \item Public blockchain networks (aka permissionless) such as Bitcoin \cite{Nakamoto2008} and Ethereum \cite{Wood2014} provide a platform for any participant to join the network, mine blocks, and execute transactions.
    \item Private blockchain networks (aka permissioned) such as Hyperledger Fabric \cite{Androulaki2018}, provide a platform for participants to join after gaining permission from the network administrator. It restricts the networks in a centralised manner, which goes against blockchain's characteristic of full decentralisation.
    \item Hybrid blockchain networks generally combine both private and public blockchain at the same time, under which a collection of predetermined nodes are responsible for approving blocks. Some scenarios where this architecture applies is when transactions are not made public by default; however, they are verifiable when needed.
\end{itemize}

\subsection{The Need for Modelling and Simulation}
Modelling and simulation are useful for analysing and predicting the performance of complex systems. Simulators capture key properties of such systems, mimic their behaviour, and allow for experimenting them without being actually implemented \cite{Haverkort1998}. Blockchain systems are typically complex \cite{ferretti2020ethereum} and composed of five layers \cite{belotti2019vademecum} \cite{zhu2019applications}: network layer, consensus layer, data layer, execution layer, and application layer. Therefore, it can be a challenging task to experiment with real-world blockchain systems and evaluate their performance. Hence, blockchain simulation is often a suitable alternative for two main reasons. First, it alleviates the burden of computing resources and financial costs needed for deploying and experimenting with blockchain systems. Second, it enables evaluating the blockchain performance for different scenarios under various parameter configurations.
\section{Related Works}
\label{related}
Anilkumar et al. \cite{Anilkumar2019} cover the popular and noteworthy blockchain-based platforms and their characteristics, such as Ethereum, IBM OBC, Intel Sawtooth Lake, BlockStream Sidechain Elements and Eris. A number of simulation platforms for Ethereum are also listed. Despite being an informative review, several aspects are missing including the recent simulators, evaluation metrics and the set of configured inputs. More recently, Wan et al. \cite{Wan2020BlockchainReview} present a review article that sheds light on the main blockchain's design principles and frontier operations. They also provide a detailed comparison between the different types of blockchain networks (public, private), as well as a brief description of proof of work (PoW) and Proof of Stake (PoS) on the consensus layer. While they extend their review coverage to include different simulation models (discrete event, stochastic, etc.), a comprehensive discussion on blockchain simulation is lacking.

To cover such a gap, Smetanin et al. \cite{Smetanin2020Evaluation} review the state-of-the-art evaluation approaches for blockchain systems with a focus on mathematical and stochastic models. Then, they discuss a set of simulators, and classify them with regard to the employed modelling approach. They also observe the lack of adoption and the lack of standardization amongst existing simulators. Although they provide coverage on common evaluation parameters and metrics of blockchain in general, they are not concerned with whether existing blockchain simulators support them. Moreover, their work does not account for blockchain simulators that already exist in the literature such as \textit{BlockSim}, \cite{Alharby2019}, eVIBES \cite{Deshpande2018}, and Shadow-Bitcoin \cite{Miller2015}, or those that have recently emerged such as  SIMBA \cite{Fattahi2020}, BlockPerf \cite{Polge2021}, and BlockEval  \cite{Gouda2021}. The authors of \cite{Smetanin2020Modeling} follow similar research conduct; however, they generalise their review to include other Distributed Ledger Technologies (DLT) such as Block-based Directed Acyclic Graphs (BlockDAG) and Transaction-based Directed Acyclic Graphs (TDAG).

Paulavicius et al. \cite{Paulavicius2021} survey several blockchain simulators regardless of whether they are based on an academic research effort or published as open-source projects. They classify them based on their modelling approach, supported language/framework, an associated blockchain platform, covered layers, and availability of source code. However, they only compare a selection of them in terms of supported configuration parameters (inputs) and produced metrics (outputs).

To the best of our knowledge, there is no extensive survey given so far that accounts for the following:
\begin{itemize}
    \item A systematic mapping review that accounts for all the to-date existing blockchain simulators that are backed by a published academic research effort.
    \item A coverage of which metrics supported by existing blockchain simulators are scientifically validated/evaluated.
\end{itemize}

\section{Research Methodology}
\label{methodology}
\begin{figure}[th]
\centering
\begin{tikzpicture}[ ->,>=stealth', shorten >=1pt, auto, scale=0.45, thick, every node/.style={scale=0.45, font=\sffamily\normalsize,  rounded rectangle, node distance = 50}]  
   \node (1) [minimum width=4.0cm, minimum height=1.2cm, align=center, draw] {Definition of\\Research Question};
   \node [above  of = 1, node distance = 30] {Process Steps};
   \node (2) [minimum width=4.0cm, minimum height=1.2cm, below of = 1, draw] {Review Scope};
   \node [below  of = 2, node distance = 30] {Outcomes};
   \node (3) [minimum width=4.0cm, minimum height=1.2cm, right of = 1, node distance = 115, draw] {Conduct Search};
   \node (4) [minimum width=4.0cm, minimum height=1.2cm, below of = 3, draw] {All Papers};
   \node (5) [minimum width=4.0cm, minimum height=1.2cm, right of = 3, node distance = 115, draw] {Screening of Papers};
   \node (6) [minimum width=4.0cm, minimum height=1.2cm, below of = 5, draw] {Relevant Papers};
   \node (7) [minimum width=4.0cm, minimum height=1.2cm, right of = 5, node distance = 115, align=center, draw] {Keywording using\\Abstracts};
   \node (8) [minimum width=4.0cm, minimum height=1.2cm, below of = 7, align=center, draw] {Classification\\Scheme};
   \node (9) [minimum width=4.0cm, minimum height=1.2cm, right of = 7, node distance = 115, align=center, draw] {Data Extraction\\and Mapping Process};
   \node (10) [minimum width=4.0cm, minimum height=1.2cm, below of = 9, draw] {Systematic Map};

   \path[every node/.style={font=\sffamily\small}] 
   (1) edge node [left] {} (2)
   (2) edge node [left] {} (3)
   (3) edge node [left] {} (4)
   (4) edge node [left] {} (5)
   (5) edge node [left] {} (6)
   (6) edge node [left] {} (7)
   (7) edge node [left] {} (8)
   (8) edge node [left] {} (9)
   (9) edge node [left] {} (10);
          
\end{tikzpicture}
	\caption{Steps of the systematic mapping study.}
	\label{steps}
\end{figure}
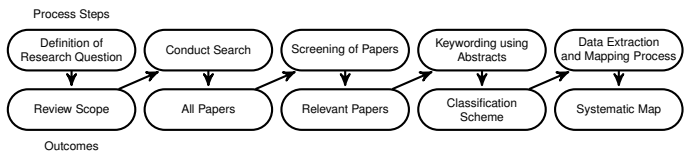

This paper conducts a systematic mapping study \cite{Petersen2008} with the aim to investigate studies pertaining to blockchain simulators.The reason for adopting a systematic mapping method in this paper is to go beyond the shallow description of existing blockchain simulators. That is, a systematic mapping review does not only help narrow down the subject exploration to specific questions but also provides analytical methods that critically examine the literature of blockchain simulators. The findings of this study will also enable us to identify and map important research directions. Figure \ref{steps} shows the five steps of systematic mapping used in this study.
\subsection{Research Questions}
This paper aims to answer the following questions:

\textbf{RQ1. How is blockchain being simulated today in the literature?}

\textbf{RQ2. Which metrics supported by existing blockchain simulators are scientifically validated/evaluated?}

\textbf{RQ3. What are the limitations of the current simulators?}    
\subsection{Performing the Literature Search} 
\label{sec:PerformingTheLiteratureSearch}
In this stage, the recent scientific papers and articles relevant to the research topic (blockchain simulators) are identified. For this purpose, the term ``blockchain simulator" is used as the keyword to search scientific databases. To specify the search, the query execution ensures the existence of both ``blockchain" and ``simulator" in the title or abstract. Furthermore, four highly reputable scientific databases were chosen: ACM Digital Library, IEEE Explore, Springer, and Scopus. For precise, accurate, and up-to-date results, high-quality articles published in books, journals, conferences, symposiums, and workshops were selected.
\subsection{Searching for Relevant Studies}
In this stage, studies related to our research questions were searched. We utilised the same searching strategy as in \cite{AlharbySM2018}. Specifically, we eliminated all the papers that were irrelevant to the topic based on their titles. If we were uncertain about a paper, we skimmed its abstract. Generally, we utilised an exclusion criteria to filter out the search results, by which non-English papers, grey literature or newsletters, and papers with no full-text were eliminated from the search.
\subsection{Searching Abstracts for Keywords}
In this stage, keywords were used to classify the relevant papers. We utilised the same keyword process as in \cite{Paulavicius2021}. Firstly, we read the abstract of each paper to spot the most significant keywords and their primary contributions. Secondly, we used these extracted keywords to classify the paper. Once all papers were classified, the papers were investigated and, if needed, switches between classification were made.
\subsection{Data Extraction and Mapping Processes}
In this stage, the required information was gathered from the papers according to their relevance to the above-stated research questions. Thus, we gathered different items of data from each study, which, in turn, highlighted the objectives and contributions of the studies.
\section{Study Results}
\label{result}
This section is designed mainly to outline the results of the systematic mapping study carried out on blockchain simulators. The results of searching and screening for relevant papers are discussed. Afterwards, the resulting classification is given.
\subsection{Searching and Screening Results}
As discussed above in Section \ref{methodology}, searching and screening are two steps in the systematic mapping study. In the searching phase, we searched for all papers using the keyword `blockchain simulator' in different scientific databases as stated above. The search returned 259 papers in total (as of 7 January 2022). In the screening phase, upon investigating the title and abstract of the papers, we excluded 209 irrelevant papers. These excluded papers are those whose main contribution is not focused on simulating the blockchain. The reason behind the high number of eliminated papers is twofold. Firstly, many papers were irrelevant to our study, since our focus was to explore blockchain simulators from a technical perspective. Secondly, some of the excluded papers were about general aspects of blockchains, with no contributions related to our pre-defined research questions. After that, duplicate papers, specifically 23, were removed, resulting in 27 final papers. Finally, we excluded 7 papers that were relevant to specific applications; i.e. they provided no useful information on simulation for general blockchains. As a consequence, we ended this phase with 20 papers on which to carry out our systematic mapping study.
\subsection{Taxonomy for Blockchain Simulators}
Following the \textit{keywording strategy} discussed in Section \ref{methodology}, we characterised the simulators according to several criteria:

\begin{enumerate}
    \item \textbf{Overview of blockchain simulators}: This represents further information related to the identified blockchain simulators in Table \ref{summary}.
    \item \textbf{Comparative analysis}: This critically compares existing simulators based on their available open-source implementation. The comparison is focused on their supported configuration parameters (inputs) and generated metrics (outputs) as per Table \ref{parTable}.
    \item \textbf{Code Quality}: This provides a general report about the source code including number of bugs, code smell, and security hotspot as shown in Table \ref{codeQuality}.
    \item \textbf{Scientifically Validated/Evaluated Metrics}: This is to represent which supported metrics by each simulator are scientifically validated/evaluated in their corresponding papers, as per shown in Table \ref{evalMet}.
\end{enumerate}

\subsection{Overview of Blockchain Simulators}
The systematic review resulted in 20 relevant blockchain simulators. Table \ref{summary} lists and examines them based on the following criteria:
\begin{enumerate}
    \item Code availability: reflects if the source code of the simulator is publicly available on GitHub.
    \item Programming language and library: reflects the programming language and the libraries used for coding the simulator.
    \item Core of the simulator: reflects if the simulator inherits a base simulator or is built from scratch.
    \item Purpose and objective: reflects if the simulator is designed for assessing performance and/or security.
    \item Blockchain platform: reflects the type/platform of the simulated blockchain; i.e. Bitcoin, Ethereum, and IOTA.
    \item Consensus algorithm: reflects the implemented consensus algorithm in the simulator.
\end{enumerate}

This section provides a summary of each blockchain simulator as follows. An early attempt in 2015 was carried out by Miller and Jansen \cite{Miller2015} who proposed a \textit{discrete event} Shadow-Bitcoin simulator. Its main focus was to simulate Bitcoin networks. This simulator utilises the concept of shadowing which permits using a parallel processing concept. Accordingly, the simulator has the ability to provide insights about the simulated multi-threading application in a scalable manner. Shadow-Bitcoin simulator is recognised as a dynamic and stochastic simulator developed using pure Python. The code is publicly available at GitHub. PoW consensus algorithm is implemented in this simulator. The simulator is able to focus on transaction propagation and providing insights about the system's performance and security.

Wang and Kin \cite{Wang2020} propose a blockchain simulator named FastChain, which extends Shadow-Bitcoin to support evaluating the correlation between throughput, block propagation time, and bandwidth-informed neighbor selection algorithms. FastChain facilitates the tuning of several parameters that influence the blockchain's performance with regard to block rate and throughput.

A similar attempt is carried out by Stoykov et al. \cite{Stoykov2017}, who proposed a \textit{discrete-event} and dynamic simulator named \textit{VIBES}, which stands for Visualisations of Interactive, Blockchain, Extended Simulations. It is coded in Scala with the aim of enabling empirical insights and analytics about the blockchain performance under various parameters, such as network topology and area size, are simulated, with the aims of predicting the total processing time, the total number of transactions processed, throughput (transactions per second), block propagation delay. While VIBES focuses on simulating Bitcoin-style blockchain networks, Ethereum is out of its scope. Following this, Deshpande et al. \cite{Deshpande2018} proposed \textit{ethernet VIBES} (eVIBES), a further improvement of VIBES, to mimic the behaviour of the Ethereum network. It depends on a reactive manifesto model using an orchestrator and reducer in its core. The orchestrator is used to control the simulator in which it receives the parameter settings from the user and is fed to Ethereum network.

Another \textit{discrete-event}, dynamic and stochastic simulator for simulating generic blockchain is proposed by Piriou and Dumas \cite{Piriou2018}. The authors find that the previous simulators depend on applying consensus algorithms in a sequential manner which may result in a double-spending attack. Accordingly, they mainly contribute a generic blockchain-style simulator with a focus on distributed consensus protocols; namely, PoW and PoS. The source code is written in Python and utilises PyCATSHOO library, which allows dealing with a large number of parameters of interest. The simulator sheds light on the impact of various parameters on the overall performance. This is done through the integration with the Monte Carlo simulation, which is known for its ability to check the dynamic behavior.

Wang et al. \cite{Wang2018} also proposed a \textit{discrete-event} and stochastic simulator that has the ability to simulate the complex and dynamic behaviour of Bitcoin blockchain network. The main aim of this simulator is to evaluate the blockchain performance by setting various parameters such as simulation time, number of nodes, mining time, block size and transaction size.

Aside from traditional simulations, Memon et al. \cite{Memon2018} proposed a \textit{queuing} blockchain simulator using the M/M/n/L queuing system. This simulator is coded in Java and was designed with the aim of simulating PoW-based mining operations, which are known to be costly and time-consuming tasks.

\textit{BlockSim} is one of the well-established simulators, which is initially proposed by Alharby and Moorsel \cite{Alharby2019}; and has been further developed in \cite{Alharby2020}. The source code, written in pure Python, is publicly available on GitHub. This simulator aims to mimic the implementations of public blockchains (Bitcoin and Etherum) using PoW. Like other \textit{discrete-event} simulators, it enables testing the influence of various configurations on the overall blockchain's performance. This is done via two different modes of simulation which are \textit{full transactions} and \textit{light transactions} techniques. The former emulates a realistic blockchain network and records detailed logs of a typical transaction journey. According to the simulator's authors, this technique is time and resources intensive; however, provides an in-depth insight into the network latency measurement. On the other hand, the latter simulates the blockchain network's behaviour by employing a single transaction pool, and omits several transaction details, which is, according to its authors, more effective in terms of time and computing consumption; however, it can be useful for other purposes other than latency measurement such as transactions throughput and execution cost. 

Further, Polge et al. \cite{Polge2021} appraise \textit{BlockSim} \cite{Alharby2019} simulator's performance with Bitcoin. However, as per their study, most of the existing simulators lack several important features such as extensibility and failure in covering all aspects/metrics. Therefore, an extended version referred to as BlockPerf is proposed to alleviate BlockSim issues. It is also written in Python with the aim to realise a stochastic, dynamic and \textit{discrete event} simulator depending on PoW consensus protocol. Similarly, Agrawal et al. \cite{Agrawal2020} state that \textit{BlockSim} \cite{Alharby2019} is restricted to simulating blockchain networks (either Bitcoin or Ethereum) over a single CPU which results in bottleneck problems. Therefore, \textit{BlockSim-Net}, another extended version of BlockSim, is proposed as a distributed simulator. What distinguishes BlockSim-net from the traditional \textit{BlockSim} is the ability to focus on the actual propagation of blocks. Furthermore, it is useful for the assessment of blockchain application security (such as a selfish mining attack on PoW). 

Another \textit{discrete-event} blockchain simulator is proposed by Faria and Correia \cite{Faria2019}; also named as BlockSim. Not to be confused with other \textit{BlockSim} simulator in \cite{Alharby2019}, it is also coded in Python and leverages the SimPy 3 library. Unlike the other \textit{BlockSim}, it simulates blockchain networks over specific intervals. Thus making it a \textit{discrete-event}, stochastic and a \textit{dynamic} simulator. 

Fattahi et al. \cite{Fattahi2020} have stated that BlockSim, proposed by Faria and Correia \cite{Faria2019}, is a reliable simulator able to evaluate blockchains. However, it does not simulate some real features, such as Merkle tree transactions. Therefore, they proposed an extended version of BlockSim, referred to as SIMulator, for application to blockchains (SIMBA). Similar to its forerunner, it is written in Python and uses SimPy 3 features. Also similar to BlockSim \cite{Faria2019}, it is a \textit{stochastic} and \textit{discrete-event} simulator.

Pandey et al. \cite{Pandey2019} also proposed another blockchain simulator named as BlockSIM, with "sim" capitalised. BlockSIM. It is a \textit{stochastic discrete-event} and \textit{dynamic} simulator written in Python using SimPy library. It facilitates evaluating the performance of Ethereum and Hyperledger blockchain networks and supports both of the PoW and PoA protocols. More recently, in 2020 Alsahan et al. \cite{Alsahan2020} extended this work and proposed a local Bitcoin simulator that has the ability to enable fast simulation for large scale networks without affecting the mining process quality. It is a \textit{virtualization} based simulator with the ability to model different network topologies.

Wang et al. \cite{BWang2020} proposed a ChainSim simulator that evaluates peer to peer blockchain networks. The simulator's main aim is to alleviate the burden of computing resources and the financial cost needed for deploying and experimenting with blockchain systems. Such a simulator has the ability to simulate blockchain-based applications with thousands of involved nodes.

Another direction is to simulate concurrent operations within blockchain networks. To do so, Gouda et al. \cite{Gouda2021} propose the Blockeval simulator, which uses deep learning algorithms to allow simulating scalable blockchain systems. This makes Blockeval's a modular simulator for assessing the performance of private blockchain networks. Moreover, it can elaborate on the metrics used for assessing the system. Blockeval's main contribution is twofold. Firstly, it can be used to assess the scalability of the proposed blockchain system. Secondly, it can analyse the security of the proposed blockchain system. 

Another attempt that was modelled with both PoW and PoS protocols was carried out by Aoki et al. \cite{Aoki2019}. The simulator, the SimBlock was written in Java. It can be differentiated from its peers by its ability to investigate blockchain performance with different behaviours of nodes. SimBlock is considered as a \textit{stochastic, dynamic}, and discrete-event simulator that focuses on modelling block generation and message transmissions. 

This work has been further extended by Basile et al. \cite{Basile2021}. They state that SimBlock omits simulating the block mining process; thus, they address this limitation in their work. Banno and Shudo \cite{Banno2019} also extended SimBlock to support to simulate thousands of nodes and to improve the neighbor selection strategy. Moreover, it enables assessing the influence of relay networks on the overall performance. 

Beyond typical blockchain data structure, Zander et al. \cite{Zander2019} proposed the DAGsim simulator that uses a \textit{Directed Acyclic Graph} (DAG) approach to represent a scalable distributed ledger, which allows simulating a vast amount of transactions over a large number of nodes. This simulator is influenced by the philosophy of a DLT network called IoTA \cite{IoTAWebPage}. An interesting feature provided by DAGsim is the support of modelling and experimenting with malicious nodes. The code is written in Python with the ability to run in $O(n^2)$, where $n$ is the number of nodes. This made DAGsim an \textit{agent based} stochastic and dynamic simulator. The final output of the simulator is a DAG representing the structure of all transactions. 

By and large, there are 4 simulation models found in the literature of blockchain simulators, namely, \textit{stochastic, dynamic, discrete event, virtualization} models. Bear in mind that all the simulators are discrete-event except the queuing model simulation in \cite{Memon2018}, the agent-based simulation in \cite{Zander2019}, and the virtualization based modelling in \cite{Alsahan2020}. We also find that existing blockchain simulators do not support consensus mechanisms other than PoW, PoA and PoS; among them, PoW is the only protocol implemented in all simulators. As regarding PoS and PoA, there is only one simulator that implements each of them. Another interesting observation is that Bitcoin is the most popular blockchain network, being implemented in 15 out of the 21 covered simulators.

\begin{table*}[t]
\setlength{\tabcolsep}{7pt}
\renewcommand{\arraystretch}{0.60}
    \centering
	\caption{A summary of blockchain simulators. Each row represents a separate simulator, while the columns represent the features. Note that all the simulators are stochastic dynamic simulators and characterized as being discrete-event except the three simulators; namely: Modelling by queuing theory simulation \cite{Memon2018} that is queuing model, DAGsim \cite{Zander2019} that is agent-based and Local Bitcoin \cite{Alsahan2020} that is virtualization based.}
    \label{summary}
  \begin{tabular}{lcccccccccc}
    \hline 
    Simulator & 
    Year & 
    GitHub code & 
    Prog. lang. & 
    Library & 
    Core &
    PRF. &
    Security & 
    Platform &
    Consensus &
     \\

    \hline
     Shadow-Bitcoin \cite{Miller2015}     & 2015     & \cmark      & Python  & N/A     &Shadow          &\cmark  &\cmark      &Bitcoin   & PoW   \\
    \hline
    VIBES \cite{Stoykov2017}     &2017     & \cmark     & Scala  & N/A     & N/A         &\cmark & \cmark     & Bitcoin   & PoW        \\
    \hline
    Stochastic Blockchain& 2018     & N/A     & Python  & PyCATSHOO     & N/A           &\cmark & \cmark      & Generic  & PoW  \\
    models \cite{Piriou2018}   &      &     &   &      &           & &      &   &/PoS   \\
              
    \hline
     Behavior and Quality     & 2018     & N/A     & Python  & SimPy     & N/A          & \cmark & \xmark      &Bitcoin  & PoW  \\
      of Blockchain  \cite{Wang2018}     &      &    &&      &           & &      &  &   \\
         
    \hline
    eVIBES \cite{Deshpande2018}     & 2018     &\cmark     &Scala  & N/A     &\cite{Stoykov2017}         &\cmark  &\xmark     &Ethereum   & PoW \\   
    \hline
    Modeling by Queuing      & 2018     & N/A     & Java  & N/A     & \cite{Bertoli2009}        &\cmark & \xmark      &Bitcoin  & PoW         \\
    Theory \cite{Memon2018}         &     &   &      &     &      &      &  &     &       \\
    \hline
    BlockSIM \cite{Pandey2019}     & 2019     &\cmark      &Python  & SimPy 3.0     &N/A         &&\xmark     &Ethereum/   & PoW     \\
          &      &      &  &     &   &      &     &Hyperledger    & /PoA           \\
    \hline
    DAGsim \cite{Zander2019}     & 2019     &N/A      &Python  & N/A     &N/A       &\cmark&\xmark     &IOTA   & IOTA       \\
    \hline
    \textit{BlockSim} \cite{Alharby2019}     & 2019     &\cmark      &Python  & N/A     &N/A         &\cmark&\xmark     &Bitcoin/   &PoW      \\
            &      &  &     &     &   &  &     &Ethereum    &       \\
      \hline
    FastChain \cite{Wang2020}     & 2019     &N/A      &N/A  & N/A     &\cite{Miller2015}    &\cmark &\xmark     &Bitcoin   &Pow           \\
    \hline
    simBlock \cite{Aoki2019}     & 2019     &\cmark      &Java  & N/A     &N/A         &\cmark &\xmark     &Bitcoin   &PoW \\
    &      &      &  &      &         &&     &   &/PoS \\
    
    \hline
    Blocksim \cite{Faria2019}     & 2019     &\cmark       &Python  & SimPy 3.0     &N/A        &\cmark &\xmark     &Bitcoin/   &PoW          \\
    &      &      &  &     &     &        &  &Ethereum    &        \\
     \hline
    \textit{Ext-} simblock \cite{Banno2019}     & 2019     &N/A       &Java  & N/A     &\cite{Aoki2019}      &\cmark&\xmark     &Bitcoin   &PoW          \\
    \hline
    BlockSim-Net \cite{Agrawal2020}     & 2020     &N/A       &Python  & N/A     &\cite{Alharby2019}          &\cmark&\xmark     &Bitcoin/   &PoW         \\
     &      &      &  &     &        & &    &Ethereum    &        \\
     \hline
    ChainSim \cite{BWang2020}     & 2020     &N/A       &Python  & N/A     &\cite{Alharby2019}       & \cmark &\xmark     &Bitcoin   &PoW           \\
    &      &      &  &          &    & &     &Ethereum    &\\
     \hline
    Local Bitcoin      & 2020     &\cmark      &Python  & N/A     &N/A      &\cmark  &\xmark     &Bitcoin   &PoW           \\
    Network  \cite{Alsahan2020}&            &  &     &     &     &     &    &          \\
    \hline
    SIMBA \cite{Fattahi2020}      & 2020     &\cmark       &Python  & SimPy 3.0     &\cite{Faria2019}       &\cmark &\xmark     &Bitcoin   &PoW          \\
    \hline
    \textit{Ext 2 -} simBlock  \cite{Basile2021}      & 2021     &N/A       &Java  & N/A     &\cite{Aoki2019}        &\cmark &\xmark     &Bitcoin   &PoW          \\
    \hline
    BlockPerf  \cite{Polge2021}      & 2021     &\cmark       &Python  & N/A     &\cite{Alharby2019}     & \cmark  &\xmark     &Bitcoin   &PoW           \\
    \hline
    BlockEval  \cite{Gouda2021}      & 2021     &\cmark       &Python  & SimPy     &N/A        &\cmark &\xmark     &Bitcoin   &PoW     \\

    \hline
  \end{tabular}
  \end{table*}
	
\subsection{Comparative Analysis}
\label{parameter}
After searching the internet for blockchain simulators and noting their main design principles, we focused on the operational range of each simulator. Specifically, we study the set of supported configuration parameters (inputs) and provided metrics (outputs) by each simulator.
A brief description of each is given below.
\begin{table*}[]
    \centering
    \caption{The definition of the parameters and metrics used with respect to blockchain layers.
    }
    \label{definitions}
    \begin{tabular}{p{1cm}p{3.7cm}p{3.6cm}p{3.7cm}p{3.6cm}}
\toprule 
 \multicolumn{1}{l}{{\footnotesize \textbf{Layer}}} & \multicolumn{1}{c}{{\footnotesize \textbf{P/M}}} & \multicolumn{1}{c}{{\footnotesize \textbf{Definition}}} & \multicolumn{1}{c}{{\footnotesize \textbf{P/M}}} & \multicolumn{1}{c}{{\footnotesize \textbf{Definition}}} \\
\midrule 
 \multicolumn{1}{c}{\multirow{7}{*}{{\footnotesize {Network Layer}}}} & {\footnotesize \textbf{(P1)} Total number of nodes} & {\footnotesize number of involved nodes} & {\footnotesize \textbf{(P7)} Payload transaction size (unit: Megabyte) } & {\footnotesize maximum transaction size} \\
  & {\footnotesize \textbf{(P1.1)} Regions of nodes (unit: geographical)} & {\footnotesize  geographical location of each node.} & {\footnotesize \textbf{(P8)} Block size (unit: Megabyte) } & {\footnotesize maximum block size configured} \\
  & {\footnotesize \textbf{(P2)} Total number of connections } & {\footnotesize number of possible connections} & {\footnotesize \textbf{(M1)} Average block size} & {\footnotesize average block sizes} \\
  & {\footnotesize \textbf{(P3)} Average block propagation delay (unit: seconds)} & {\footnotesize average time delayed in the propagation process of each block} & {\footnotesize \textbf{(M2)} Average block propagation time} & {\footnotesize average time taken to propagate blocks} \\
  & {\footnotesize \textbf{(P4)} Average transaction propagation delay (unit: seconds) } & {\footnotesize verage time delayed in the propagation process of each transaction} & {\footnotesize \textbf{(M3)} Average transaction propagation time} & {\footnotesize average time taken by the simulator to propagate transaction} \\
  & {\footnotesize \textbf{(P5)} Average bandwidth (unit: bits per second) } & {\footnotesize bandwidth assumed for the simulated network} & {{\footnotesize \textbf{(M4)} Throughput (unit: Tx/second)}} & {{\footnotesize throughput taken to the end of the simulations.}} \\
  & {\footnotesize \textbf{(P6)} Average latency \ (unit: seconds)} & {\footnotesize average latency assumed for the simulated network} &   &   \\
\hline\hline 
 \multicolumn{1}{l}{\multirow{2}{*}{{\footnotesize  {Data layer}}}} & {\footnotesize \textbf{(P9)} Generate random transactions (unit: Integer of Tx per second)} & {\footnotesize automatically generated transactions} & \multirow{2}{*}{{\footnotesize \textbf{(M6)} Security}} & \multirow{2}{*}{{\footnotesize  security assessement}} \\
  & {\footnotesize (\textbf{M5)} Chain of block} & {\footnotesize resulting chain} &   &   \\
\hline\hline 
 \multicolumn{1}{l}{\multirow{5}{*}{{\footnotesize {Consensus layer}}}} & {\footnotesize \textbf{(P10)} Average mining power (Hash Rate)} & {\footnotesize average used mining power } & {\footnotesize \textbf{(M7)} Average block interva} & {\footnotesize average time for the blocks to accept transactions} \\
  & {\footnotesize \textbf{(P11)} PoW consensus Algorithm } & {\footnotesize ability to implement PoW consensus algorithm.} & {\footnotesize \textbf{(M8)} Number of generated blocks} & {\footnotesize  total number of generated blocks} \\
  & {\footnotesize \textbf{(P12)} Other consensus Algorithm} & {\footnotesize ability to implement other consensus algorithms than PoW.} & {\footnotesize \textbf{(M9)} Number of mined blocks} & {\footnotesize total number of mined blocks } \\
  & {\footnotesize \textbf{(P13)} Average transaction fee (unit: cryptocurrency)} & {\footnotesize average transaction fees} & {\footnotesize \textbf{(M10)} Rate of orphan blocks (unit: percentage)} & {\footnotesize percentage of the orphan blocks} \\
  & {\footnotesize \textbf{(P14)} Block Interval (unit: seconds)} & {\footnotesize Average time for creating a block} & {\footnotesize \textbf{(M11)} Fork Resolution} & {\footnotesize determine forks occurred as protocol change.} \\
\hline\hline 
 \multicolumn{1}{l}{{\footnotesize {Incentive layer}}} & {\footnotesize \textbf{(P15)} Reward for mining a new block (unit: cryptocurrency)} & {\footnotesize amount of reward configured} & {\footnotesize \textbf{(M12)} Reward for miner (unit: cryptocurrency)} & {\footnotesize amount of reward consumed } \\
\hline\hline 
 \multicolumn{1}{l}{{\footnotesize {Execution layer}}} & {\footnotesize \textbf{(M13) }Time of executing a contract (unit: seconds)} & {\footnotesize time taken to execute a contract.} & {\footnotesize \textbf{(M14)} Validation of contract and execution time} & {\footnotesize how the simulator validates the contract} \\
\hline\hline 
 \multicolumn{1}{l}{\multirow{2}{*}{{\footnotesize  {Application layer}}}} & {\footnotesize \textbf{(P16)} Simulation run time (unit: seconds)} & {\footnotesize configured simulation time } & {{\footnotesize \textbf{(M16)} Simulation time (unit: seconds)}} & {{\footnotesize represents the actual time taken by the simulator}} \\
  & {\footnotesize \textbf{(M15)} Resources usage } & {\footnotesize how the simulator keep track of the resource usage/utilization} &   &   \\
 \bottomrule
\end{tabular}
\end{table*}

To this end, no existing blockchain simulator can support all configurations parameters (P) and produced metrics (M). Table \ref{definitions} highlights these parameters/metrics and their association to each of the blockchain layers. By inspecting Table \ref{parTable}, we can notice that out of the 16 parameters, the least number of implemented parameters is 5, which is the case with Local Bitcoin \cite{Alsahan2020}. In other words, at least 40\% of the parameters are implemented. On the other hand, at most about 81\% of the parameters are implemented, which is the case with both BlockSim \cite{Faria2019} and BlockPerf \cite{Polge2021}; i.e. 13 parameters. Similarly, not all the metrics are supported by all simulators. The least number of supported metrics is 4, which represents 25\% of the metrics as is with BlockSIM \cite{Pandey2019}. By contrast, at most 88\% of the parameters are implemented, which is the case with BlockPerf \cite{Polge2021}; i.e. 13 parameters. Accordingly, we can notice that BlockPerf\cite{Polge2021} is the richest simulator with both parameters and metrics.

\begin{table*}[t]
\renewcommand{\arraystretch}{0.8}
    \centering
	\caption{Set of parameters available in each simulator. For a detailed description of the parameters refer to Subsection \ref{parameter}. The sign $\CIRCLE$  means that the parameter is  available in the simulator, while the sign  $\Circle$  means that the parameter is not available in the simulator. The last row represents the total number of simulators supporting particular parameter. Similarly, the last column represents the total number of parameters supported by a particular simulator. The bold values represent the maximum values and the underlined values represent the minimum values.}
    \label{parTable}
  \begin{tabular}{lccccccccccccccccc||c}
    \hline 
    Simulator&\multicolumn{16}{c}{\textbf{Parameters}}\\\cline{2-19}
    &P1&P1.1 &P2 &P3&P4&P5&P6 &P7&P8&P9&P10&P11&P12&P13&P14&P15&P16&\textbf{Total}
    \\\hline
     Shadow-Bitcoin \cite{Miller2015} & \CIRCLE& \CIRCLE&\Circle &\Circle &\Circle &\CIRCLE &\Circle &\Circle &\Circle &\CIRCLE &\Circle &\CIRCLE&\Circle &\Circle &\Circle &\CIRCLE &\Circle&6   \\
     \hline
     VIBES \cite{Stoykov2017} & \CIRCLE& \Circle&\CIRCLE &\CIRCLE &\CIRCLE &\Circle &\Circle &\CIRCLE &\CIRCLE &\CIRCLE &\CIRCLE &\CIRCLE &\Circle&\Circle &\Circle &\CIRCLE &\Circle&10    \\
     \hline
     eVIBES \cite{Deshpande2018} & \CIRCLE& \Circle&\Circle &\Circle &\Circle &\Circle &\Circle &\CIRCLE &\Circle &\CIRCLE &\CIRCLE &\CIRCLE &\Circle&\Circle &\Circle &\CIRCLE &\CIRCLE&6    \\
     \hline
     BlockSIM \cite{Pandey2019} & \CIRCLE& \Circle&\Circle &\Circle &\Circle &\Circle &\Circle &\CIRCLE &\CIRCLE &\CIRCLE &\CIRCLE &\CIRCLE &\Circle &\Circle &\CIRCLE &\Circle &\CIRCLE&8    \\
     \hline
     \textit{BlockSim} \cite{Alharby2019} & \CIRCLE&\Circle&\Circle &\CIRCLE &\CIRCLE &\Circle &\Circle &\CIRCLE &\CIRCLE &\CIRCLE &\CIRCLE &\CIRCLE &\Circle &\CIRCLE &\CIRCLE &\CIRCLE &\CIRCLE&12   \\
     \hline
     SimBlock \cite{Aoki2019} & \CIRCLE& \CIRCLE&\CIRCLE &\Circle &\Circle &\CIRCLE &\CIRCLE &\Circle &\CIRCLE &\Circle &\CIRCLE &\CIRCLE & \CIRCLE &\CIRCLE &\CIRCLE &\Circle &\CIRCLE&12    \\
     \hline
     BlockSim \cite{Faria2019} & \CIRCLE&\CIRCLE&\CIRCLE &\CIRCLE &\Circle &\CIRCLE &\CIRCLE &\CIRCLE &\CIRCLE &\CIRCLE &\CIRCLE &\CIRCLE & \Circle &\CIRCLE &\CIRCLE &\Circle &\Circle&\textbf{13}    \\
     \hline
     Local Bitcoin  \cite{Alsahan2020} &\CIRCLE &\Circle &\Circle &\CIRCLE &\Circle &\Circle &\Circle &\Circle &\Circle& \Circle&\CIRCLE &\CIRCLE &\Circle  &\Circle &\Circle &\Circle &\CIRCLE&\underline{5}    \\
     \hline
     SIMBA \cite{Fattahi2020} &\CIRCLE& \CIRCLE&\CIRCLE&\CIRCLE &\Circle &\CIRCLE &\CIRCLE &\CIRCLE &\CIRCLE &\CIRCLE &\CIRCLE &\CIRCLE &\Circle &\Circle &\CIRCLE &\Circle &\Circle&12    \\
     \hline
     BlockPerf \cite{Polge2021} & \CIRCLE& \CIRCLE&\Circle &\CIRCLE &\CIRCLE &\Circle &\Circle &\CIRCLE &\CIRCLE &\CIRCLE &\CIRCLE &\CIRCLE &\Circle &\CIRCLE&\CIRCLE &\CIRCLE &\CIRCLE &\textbf{13}   \\
     \hline
     BlockEval \cite{Gouda2021} & \CIRCLE& \CIRCLE&\CIRCLE&\CIRCLE&\Circle &\Circle &\Circle &\CIRCLE&\CIRCLE&\CIRCLE&\Circle &\CIRCLE & \Circle &\Circle &\CIRCLE &\CIRCLE &\CIRCLE&11    \\
     \hline \hline
    \textbf{Total}  &\textbf{11}&6&5&7&3&4&3&8&8&9&9&\textbf{11}&\underline{1}&4&7&6&7& \\
      \hline\hline 
    Simulator&\multicolumn{17}{c}{\textbf{Metrics}}\\\cline{2-19}
    &M1&M2&M3&M4&M5&M6&M7&M8&M9&M10&M11&M12&M13&M14&M15&M16&&\textbf{Total}\\

    \hline
     Shadow-Bitcoin \cite{Miller2015} & \Circle&\Circle &\Circle &\CIRCLE &\Circle &\CIRCLE &\CIRCLE &\Circle &\CIRCLE &\Circle &\Circle &\CIRCLE &\CIRCLE &\Circle &\CIRCLE &\CIRCLE& & 8    \\
     \hline
     VIBES \cite{Stoykov2017} & \CIRCLE&\CIRCLE &\CIRCLE &\CIRCLE &\CIRCLE &\CIRCLE &\CIRCLE &\CIRCLE &\CIRCLE &\CIRCLE &\Circle &\Circle &\Circle &\Circle &\Circle &\CIRCLE&&11     \\
     \hline
     eVIBES \cite{Deshpande2018} & \CIRCLE&\CIRCLE &\Circle &\CIRCLE &\Circle &\Circle &\CIRCLE &\CIRCLE &\CIRCLE &\CIRCLE &\CIRCLE &\CIRCLE &\Circle &\CIRCLE &\Circle &\CIRCLE&&11     \\
     \hline
     BlockSIM \cite{Pandey2019} &\CIRCLE&\Circle &\Circle &\CIRCLE &\Circle &\Circle &\CIRCLE &\CIRCLE &\Circle &\Circle &\Circle &\Circle &\Circle &\Circle &\Circle &\Circle&&\underline{4}     \\
     \hline
     \textit{BlockSim} \cite{Alharby2019} &\CIRCLE&\Circle &\CIRCLE &\CIRCLE &\CIRCLE &\Circle &\Circle &\CIRCLE &\CIRCLE &\CIRCLE &\CIRCLE &\CIRCLE &\Circle &\Circle &\Circle &\CIRCLE&&10     \\
     \hline
     SimBlock \cite{Aoki2019} &\CIRCLE&\CIRCLE &\Circle &\Circle &\CIRCLE &\Circle &\CIRCLE &\CIRCLE &\Circle &\CIRCLE &\Circle &\Circle &\Circle &\Circle &\Circle &\CIRCLE&&7     \\
     \hline
     BlockSim \cite{Faria2019} & \CIRCLE&\CIRCLE &\CIRCLE &\CIRCLE &\Circle &\Circle &\Circle &\CIRCLE &\CIRCLE &\Circle &\CIRCLE &\Circle &\Circle &\Circle &\Circle &\Circle&&7     \\
     \hline
     Local Bitcoin  \cite{Alsahan2020} & \Circle&\Circle &\Circle &\CIRCLE &\Circle &\Circle&\Circle &\CIRCLE &\CIRCLE &\Circle &\CIRCLE &\CIRCLE &\Circle &\Circle &\CIRCLE &\Circle&&6     \\
     \hline
     SIMBA \cite{Fattahi2020} & \CIRCLE&\CIRCLE &\Circle &\CIRCLE &\Circle &\Circle &\Circle &\CIRCLE &\CIRCLE &\Circle &\CIRCLE &\Circle &\Circle &\Circle &\Circle &\CIRCLE&&7     \\
     \hline
     BlockPerf \cite{Polge2021} & \CIRCLE&\CIRCLE &\CIRCLE &\CIRCLE &\CIRCLE &\Circle &\Circle &\CIRCLE &\CIRCLE &\CIRCLE &\CIRCLE &\CIRCLE &\CIRCLE &\CIRCLE &\CIRCLE &\CIRCLE&&\textbf{14}    \\
     \hline
     BlockEval \cite{Gouda2021} & \CIRCLE&\CIRCLE &\Circle &\Circle &\CIRCLE &\Circle &\Circle &\CIRCLE &\CIRCLE &\CIRCLE &\CIRCLE &\CIRCLE &\Circle &\Circle &\Circle &\CIRCLE&&9     \\
     \hline\hline
     Total&9&7&4&9&5&\underline{2}&5&\textbf{10}&9&6&7&6&\underline{2}&\underline{2}&3&8\\
     \hline

\end{tabular}
  \end{table*}

\subsection{Code Quality}
To solidify the view about the simulators, we assess their source code from different aspects, which will help researchers determine the future trends and modifications needed for each simulator. Below is a detailed description about the aspects used.
\begin{itemize}
    \item Lines of Code: is the number of code lines.
    \item Comments(\%): is the percentage of commented lines.
    \item Duplication(\%): is the percentage of duplicated lines.
    \item Files: is the number of code files.
    \item Bugs: is the number of bugs.
    \item Code Smells: is the complexity degree of understanding the code.
    \item Security Hotspots: is the number of source code parts that need major overhaul from the security aspect.
\end{itemize}

We have used the Sonarqube \cite{SonarQube} and the Count Lines of Code
(CLOC) \cite{adanial_cloc} tool to assess the codes of the simulators. An interesting point about such a tool is its ability to shed light on the previous aspects mentioned above. A detailed description about this comparison is shown in Table \ref{codeQuality}. With quick skimming for the Table, we can notice that eVIBES \cite{Deshpande2018} has the largest number of lines while being Bug free. On the other hand, local Bitcoin \cite{Alsahan2020} has the least number of lines while also being Bug free. Another interesting point is that 6 simulators out of 11 are Bug free. Despite being the most reputable, \textit{BlockSim} \cite{Alharby2019} has the largest number of bugs with about 6\% code duplication.
\begin{table*}[t]
\renewcommand{\arraystretch}{0.7}
    \centering
	\caption{Evaluating aspects of each simulator using Sonarqube  and CLOC tool.}
    \label{codeQuality}
    \footnotesize{
  \begin{tabular}{l *{8}{c}}
    \hline 
    Simulator& \multicolumn{7}{c}{Evaluation aspect}\\\cline{2-8}
     &Lines of Code &Comments(\%) &Duplication(\%)&Files&Bugs&Code Smells&Security Hotspots
    \\

\hline
     Shadow-Bitcoin \cite{Miller2015} & 1218 & 12 & 0 & 17 & 4 & 117 & 12     \\
     \hline
     VIBES \cite{Stoykov2017} & 20773 & 1.8 & 0 & 118 & 2 & 19 & 2    \\
     \hline
     eVIBES \cite{Deshpande2018} & 25909 & 5 & 0 & 166 & 0 & 53 & 0     \\
     \hline
     BlockSIM \cite{Pandey2019} & 712 & 22 & 0 & 35 & 0 & 28 & 4    \\
     \hline
     \textit{BlockSim} \cite{Alharby2019} & 1730 & 18 & 5.8 & 2.9 & 65 & 215 & 11     \\
     \hline
     SimBlock \cite{Aoki2019} & 2487 & 53 & 2.4 & 37 & 6 & 128 & 9    \\
     \hline
     BlockSim \cite{Faria2019}  & 1721 & 18 & 0 & 27 & 0 & 28 & 4    \\
     \hline
     Local Bitcoin \cite{Alsahan2020} & 323 & 13 & 15 & 6 & 0 & 4 & 36      \\
     \hline
     SIMBA \cite{Fattahi2020} & 2284 & 15 & 3 & 36 & 0 & 56 & 4     \\
     \hline
     BlockPerf \cite{Polge2021} & 1668 & 4 & 0 & 26 & 0 & 26 & 6     \\
     \hline
     BlockEval \cite{Gouda2021}  & 2761 & 76 & 0 & 20 & 1 & 71 & 1    \\
     \hline
\end{tabular}
}
	\end{table*}
\subsection{Scientifically Validated/Evaluated Metrics}
For the picture to be complete, we also shed light on the Scientifically validated/evaluated metrics in the corresponding paper of each simulator. A summary of this is shown in Table \ref{evalMet}. According to the reviewed simulators, there are 15 validation/evaluation metrics; a brief description of each is given below. 
\begin{itemize}

    \item Block propagation time: which is the time taken from sending to receiving a block.
    
    \item Transaction propagation time: which is the time taken from sending to receiving a transaction.
    
    \item Actual block size: which is the average block size generated during the simulation period.
    
    \item Transaction throughput: which is the rate at which a set of valid committed transactions in a defined time period.
    
    \item Network delay: which is the total delay in the network.
    
    \item Number of generated block: which is the total number of the generated blocks during a simulation period.
    
    \item Number of valid block: which is the total number of valid blocks created during the simulation.
    
    \item Block verification time: which is the average time taken to verify the generated blocks during the simulation period.
    
    \item Uncle or stale blocks: which is the number of generated uncle/stale blocks during a simulation period.
    
    \item Fork resolution: which reflects whether a fork has ever occurred or not during the simulation.
    
     \item Pending transaction: which is the simulator's ability to keep track of the transaction number while awaiting confirmation.
    
    \item Mining difficulty: which is the amount of difficulty assigned for each transaction.
    
    \item Mining reward: which is the amount of the used rewards for the mining processes occurred throughout the simulation.
    
    \item Processing speed: which is the average time taken by the simulator to carry out a specific task during the simulation period.
    
    \item System stability: Which is concerned with eventual consistency over time among participating nodes in the blockchain network in terms of the ledger replicas.
\end{itemize}
Having mentioned the previous metrics, we also focused on the distribution of these metrics over the blockchain layers. Table~\ref{evalMet} shows the set of metrics associated to each layer. Moreover, from the Table we can see for each simulator which metrics are implemented. On closer inspection, we notice that the majority of the metrics  (7 out of 15) are focusing on the consensus layer. On the other hand, the incentive layer has the least share of the metrics (only 1 metric). Also, there are 9 simulators focusing on validating and evaluating the network layer. On the consensus layer, we can notice that only 7 simulators validate such a layer. With the least attention, the incentive layer is validated in one simulator only. From another viewpoint, with  6 validation/evaluation metrics, VIBES \cite{Stoykov2017} comes out at the top whereas shadow-Bitcoin \cite{Miller2015} has the least number of validation/evaluation metrics (1~metric).

\begin{table*}[t]
\renewcommand{\arraystretch}{0.8}
    \centering
	\caption{The scientifically validated/evaluated metrics of each simulator with respect to different layers. The signs $\CIRCLE$ and $\Circle$ depicts the available and missing metrics, respectively. The last row represents the total number of simulators used particular metric. Similarly, the last column represents the total number of metrics used by a particular simulator. The bold values represent the maximum values and the underlined values represent the minimum values.}
    \label{evalMet}
  \begin{tabular}{lccccc|ccccccc|c|cc||c}
    \hline 
    Simulator& \multicolumn{15}{c}{Approach used }\\
    \cline{2-17}
     &\multicolumn{5}{c}{Network layer}  &  \multicolumn{7}{c}{Consensus layer}& 
     \multicolumn{1}{c}{Incentive layer} &
     \multicolumn{2}{c}{General}&
     \multicolumn{1}{c}{Total}
     \\\cline{2-17}
     &\spheading{Block propagation time}&
     \spheading{Transaction propagation time}&
     \spheading{Average of block size}&
     \spheading{Throughput}&
     \spheading{Network delay}&
     \spheading{Number of generated block}&
     \spheading{Number of mined block}&
     \spheading{Block verification time}&
     \spheading{The rate of stale (orphan) blocks}&
     \spheading{Fork resolution}&
     \spheading{Pending transaction}&
     \spheading{Mining difficulty}&
     \spheading{Mining reward} &
    \spheading{Processing Speed}&
    \spheading{System stability}
     \\
\hline
 Shadow-Bitcoin \cite{Miller2015}&\Circle&\CIRCLE&\Circle&\Circle&\Circle&\Circle&\Circle&\Circle&\Circle&\Circle&\Circle&\Circle&\Circle&\Circle&\Circle&\underline{1}  \\   
 VIBES \cite{Stoykov2017}&\CIRCLE&\Circle&\CIRCLE&\CIRCLE&\Circle&\CIRCLE&\Circle&\Circle&\CIRCLE&\Circle&\Circle&\Circle&\Circle&\CIRCLE&\Circle&\textbf{6} \\  
 eVIBES \cite{Deshpande2018}&\Circle&\Circle&\Circle&\Circle&\Circle&\CIRCLE&\Circle&\CIRCLE&\Circle&\Circle&\Circle&\Circle&\Circle&\Circle&\Circle&2 \\  
 BlockSIM \cite{Pandey2019}&\Circle&\Circle&\Circle&\CIRCLE&\Circle&\Circle&\Circle&\Circle&\Circle&\Circle&\Circle&\Circle&\Circle&\Circle&\CIRCLE&2 \\  
 \textit{BlockSim} \cite{Alharby2019}&\Circle&\Circle&\Circle&\CIRCLE&\Circle&\CIRCLE&\Circle&\Circle&\CIRCLE&\Circle&\Circle&\Circle&\Circle&\Circle&\Circle &3\\  
 SimBlock \cite{Aoki2019}&\CIRCLE&\Circle&\Circle&\Circle&\Circle&\Circle&\Circle&\Circle&\Circle&\Circle&\Circle&\Circle&\Circle&\Circle&\Circle&\underline{1} \\ 
 BlockSim \cite{Faria2019}&\CIRCLE&\CIRCLE&\Circle&\Circle&\Circle&\Circle&\Circle&\Circle&\Circle&\Circle&\Circle&\Circle&\Circle&\Circle&\Circle&2 \\ 
 Local Bitcoin \cite{Alsahan2020}&\Circle&\Circle&\Circle&\Circle&\CIRCLE&\Circle&\Circle&\Circle&\Circle&\Circle&\Circle&\CIRCLE&\Circle&\Circle&\Circle&2 \\  
 SIMBA \cite{Fattahi2020}&\CIRCLE&\Circle&\CIRCLE&\Circle&\Circle&\Circle&\Circle&\CIRCLE&\Circle&\Circle&\Circle&\Circle&\Circle&\Circle&\Circle&3 \\  
 BlockPerf \cite{Polge2021}&\Circle&\Circle&\CIRCLE&\CIRCLE&\Circle&\Circle&\CIRCLE&\Circle&\CIRCLE&\Circle&\Circle&\Circle&\CIRCLE&\Circle&\Circle&5 \\  
 BlockEval \cite{Gouda2021}&\Circle&\Circle&\Circle&\Circle&\Circle&\Circle&\Circle&\Circle&\Circle&\CIRCLE&\CIRCLE&\Circle&\Circle&\Circle&\Circle&2 \\  
    \hline
    \hline
   Total & \textbf{4}&2 &3&\textbf{4}&\underline{1}&3&\underline{1}&2&3&\underline{1}&\underline{1}&\underline{1}&\underline{1}&\underline{1}&\underline{1}& \\ 

\hline
\end{tabular}
	\end{table*}
\section{Discussion}
\label{discussion}
According to the results obtained in Section \ref{result}, this section is dedicated mainly to outlining proposed solutions to the predefined research questions stated in Section \ref{methodology}.

\textbf{RQ1. How is blockchain being simulated today in the literature?}

The systematic mapping study reveals that there are 11 out of the 20 blockchain simulation studies that publicly provide their source code. The majority of existing simulators are stochastic, which is immensely complex, and require an in-depth statistical capabilities for ensuring realistic outcomes. Several existing blockchain simulators are dynamic; a feature that aligns well with the blockchain nature. For instance, dynamic simulators enable investigating the behaviour of blockchain networks given the variable number of involved miners (nodes) over time.

As regarding the configuration parameters, we notice that existing blockchain simulators vary in their interest of which configuration parameters (refer to Table \ref{parTable}). Both BlockSim \cite{Faria2019} and BlockPerf \cite{Polge2021} are the richest simulators with 13 configuration parameters supported by each. On the other hand, Local Bitcoin \cite{Alsahan2020} is the least in terms of supported parameters, which enables controlling 5 parameters at most. Interestingly, both P1 and P11 are supported by all simulators which, in turn, reflect their importance. Conversely, P12 is the least supported configuration parameter. By inspecting the generated metrics by each simulator, we find that M8 is supported by all simulators except Shadow-Bitcoin \cite{Miller2015} which, in turn, implies its significance. On the other hand, M14 is the least supported metric by only two simulators which are eVIBES \cite{Deshpande2018} and BlockPerf \cite{Polge2021}.

\textbf{RQ2. Which metrics supported by existing blockchain simulators are scientifically validated/evaluated?}

To better understand the system’s behaviour, there is a need for a set of validation/evaluation metrics to assess the overall system’s performance. Generally, blockchain systems can be judged from different viewpoints as follows.
\begin{enumerate}
\item\textit{Usability and reliability}: \textit{Is the system ready for being implemented in a real world situation?} 
This is related to assessing the deployed network. The network can be assessed using two metrics: volume of P2P traffic and packet loss. The former represents the network’s ability to perform under an elevated traffic while the latter represents the ratio of the lost packets. In view of this metric, the system is usable if it is able to exchange a large amount of traffic with the least number of lost packets. 

\item\textit{Functional testing}: \textit{Is the system able to provide promising results?} This is related to assessing blockchain itself. Blockchain can be assessed using three metrics; transaction throughput, latency, and finality time. The transaction throughput represents the amount of successfully committed transactions per second. Blockchain is successful if it is able to provide high transaction throughput especially in the case of permissionless blockchain. The latency represents the time taken for the effect of the transaction to be reflected; it should be minimal. Finally, the finality time represents the amount of time taken for the transaction to be committed. This metric is of high importance as if wrongly adjusted, it decreases the system efficiency.
\item\textit{Resource testing}: \textit{Are the involved nodes operating properly?} This is related to assessing the involved nodes. Theoretically, this is assessed using the resource metrics which represent the computational power (CPU/GPU, memory, storage capacity, connectivity, and cache ratio) of the nodes. This metric is of high importance, as low resources can incur a significant negative impact on the chain. 

The systematic mapping study reveals that there is no simulator able to assess the system’s performance from all the different viewpoints. From the source code viewpoint, existing blockchain simulators support multiple metrics. However, their corresponding papers do not validate/evaluate all of them.  Again, the problem is not with the simulators themselves, but with the target application. Focusing on the network layer,  all simulators except eVIBES \cite{Deshpande2018} and BlockEval \cite{Gouda2021} are concerned with relevant metrics. VIBES \cite{Stoykov2017} implements 3 out of the 5 network metrics. BlockSim \cite{Faria2019}, SIMBA \cite{Fattahi2020} and BlockPerf \cite{Polge2021} implement only 2 out of them.  Shadow-Bitcoin \cite{Miller2015} implements only 1 network metric.

On the consensus layer, 7 simulators implement associated metrics. VIBES \cite{Stoykov2017}, eVIBES \cite{Deshpande2018}, \textit{BlockSim} \cite{Alharby2019},  BlockPerf \cite{Polge2021}, and BlockEval \cite{Gouda2021} implement only 2 out of the consensus metrics. Local Bitcoin \cite{Alsahan2020} and SIMBA \cite{Fattahi2020} implement only 1 metric related to the consensus layer. 

On the incentive layer, the mining reward metric is only supported by BlockPerf \cite{Polge2021}. Regarding the general metrics, the processing speed metric is only implemented in VIBES \cite{Stoykov2017} while the system stability metric is only supported by BlockSIM \cite{Pandey2019}.

\end{enumerate}

\textbf{RQ3. What are the limitations of the current simulators?}

The limitations of the covered simulators can be expressed by the following viewpoints. 

\begin{enumerate}
	\item \textit{Usability}: Great headway has been made in the field of simulating blockchain, but the work is limited. The usability of such simulators is hindered by the fact that there are a large number of parameters that need to be adjusted (such as the simulation scenario and the execution environment); this necessitates a deep understanding of blockchain technology. Furthermore, most of the existing simulators require coding skills or/and knowledge of command-line interfaces. However, some effort has been made to mitigate this issue through the use of web interfaces; as in VIBES \cite{Stoykov2017}, and eVIBES \cite{Deshpande2018}.
	
	\item \textit{Availability and Scalability}: the majority of the simulators virtually run multiple blockchain nodes on a single machine, which naturally suffers from limited resources. Thus, it can be challenging to generalise the outcomes of simulated blockchain models on real-world blockchain networks (i.e. resource usage and energy consumption). Moreover, there is no focus on node behaviour under different sources of power and resources; i.e. blockchain running on custom ASIC-based computers. Additionally, to the best of our knowledge, the majority of existing simulators do not draw much attention to the consensus layer; and many of them solely focus on PoW algorithms while neglecting others, such as PoS, PoA, Raft, and others.
	
	\item \textit{Applicability}: the majority of simulators are predominantly targeted for financial applications, such as Bitcoin.
    However, there is no generalisation done outside the field of finance. The hurdle is that there is no simulation to support the integration of blockchain with other technologies. For instance, none of the covered simulators are specifically tailored for experimenting with the intersection of blockchain with other domains such as IoT, Cloud, Cybersecurity, Supply Chain, and others.
\end{enumerate}
\section{Conclusions}
\label{conclusion}
Modelling and simulation have been useful in several disciplines, and blockchain is no exception. Such practice allows for experimenting with complex systems, such as blockchain systems, with minimum cost and effort. The presented systematic study mainly investigates blockchain simulators, their features and capabilities. The results showed that there are 20 simulators dedicated to this purpose. We focused on 11 simulators whose source code is publicly available. Additionally, it highlights scientifically validated/evaluated metrics supported by each simulator.

Interestingly, most existing simulators support stochastic, dynamic and discrete event modelling approaches. We find that the majority of existing blockchain simulators support dynamic modelling, which aligns well with the nature of blockchain networks. Regarding the evaluation/validation process, not all simulators are interested in the same collection of evaluation metrics or blockchain layers. Moreover, we find that not all supported metrics by each simulator are scientifically validated/evaluated in their corresponding papers.

To date, no blockchain simulator can comprehensively cover all blockchain facets. Moreover, existing blockchain simulators have little viability for being implemented with other technologies, such as cloud and IoT. To sum up, blockchain simulation is still in its infancy stages; and there must be further research effort in this direction.

\bibliographystyle{IEEEtran}
\bibliography{IEEEabrv,References}
\end{document}